\documentstyle[prd,eqsecnum,aps]{revtex}

\newcommand{\bmit}[1]{\mbox{\boldmath $#1$}}

\begin{document}
\draft
\preprint{Submitted to Il Nuovo Cimento B}
\title{Noisy Spherical Resonant Detector of Gravitational Waves: Veto on the 
Longitudinal Part of the Signal}
\author{Enrico Montanari}
\address{Department of Physics, University of Ferrara and
INFN Sezione di Ferrara, Via Paradiso 12, I-44100 Ferrara, Italy}
\address{montanari@axpfe1.fe.infn.it}
\maketitle
\begin{abstract}
The equations of a resonant sphere in interaction with $N$ secondary 
radial oscillators (transducers) on its surface have been found in 
the context of Lagrangian formalism. It has been shown the possibility
to exert a veto against spurious events measuring the longitudinal 
component of a signal. Numerical simulations has been
performed, which take into account thermal noise between resonators 
and the sphere surface, for a particular configuration of the 
transducers. 
\end{abstract}


\section{Introduction} 

The idea of a spherical detector of gravitational waves has been 
considered almost since the early seventies~\cite{forw71}. Very 
accurate calculations have been performed~\cite{ashdre75} in order to 
use a self--gravitating object as a possible detector. Basing upon 
these results other authors~\cite{wagpai77} have solved the inverse 
problem for a free sphere in interaction with a gravitational wave. In 
doing so they have shown an omnidirectional sensitivity and the 
possibility of testing the presence of a longitudinal term in the 
gravitational signal. However a free spherical 
resonator is not a good device. In fact a set of secondary mechanical 
oscillators (transducers) is needed in order to convert the mechanical
signal into an electromagnetic one (see~\cite[Thorne, p. 330]{300}). 
In the spherical geometry almost $5$ transducers are needed to obtain 
all the informations contained in the motion of its $5$ quadrupolar 
modes. The problem of the interaction of a sphere with these secondary 
resonators has been solved in some particular configuration only in 
the nineties~\cite{johmer93,zoumic95,merjoh95}. In the configurations 
envisaged by these authors the omnidirectional sensitivity and the 
possibility of total reconstruction of the signal has been shown, in 
this way realizing a complete observatory with only one device 
(cfr~\cite{ort92a,ort92b,ort93}). However, in the calculations 
performed until now, not enough attention has been paid to the 
possibility of vetoes against spurious events (see for 
instance~\cite{lobser96}) and to the influence of thermal noise in the 
reconstruction of the signal. 
In this paper we have focus our attention upon these two problems, 
solving them explicitly for the dodecahedron arrangement of the 
transducers first described by~\cite{johmer93}. Moreover a Lagrangian 
which describes the interaction of a sphere with $N$ radial 
transducers on the surface has been written. This is important in 
order to write the most general equations involving the motion of the 
modes of the sphere and of the secondary oscillators.
In this way it will be possible to undertake a systematic study of 
other transducers's configurations, enabling spherical detectors to 
improve their capabilities in the framework of gravitational 
astronomy~\cite{lobser96,coc9596}.

\section{Lagrangian of a free elastic body interacting with a 
gravitational wave}

An elastic body in interaction with a gravitational wave could be 
described by ordinary elastic theory if its dimensions are very 
small compared to the gravitational wavelength. Gravitation acts 
as an external field of density force (e.g.~\cite{wagpai77})
\begin{equation}
f^i = \rho\,R^i = \rho\,E_{ij}(t)\,x^j
\label{gravityforce}
\end{equation} 
where $E_{ij}$ are the components of the ``electric'' part of 
the Riemann tensor $R_{i0j0}$.
If $\bmit u  (\bmit x,t)$ represents the displacement from the 
equilibrium position in the point $\bmit x$ then 
(e.g.~\cite{le})
\begin{equation}
\rho {\partial ^2 \bmit u \over{ \partial t^2}} = \mu \ 
{\bmit \nabla}^2 
\bmit u + ( \lambda + \mu)\ {\bmit \nabla} 
({\bmit \nabla} \cdot \bmit u) + \rho\,\bmit R
\label{elasteq}
\end{equation}
Using the standard approach the solution of~(\ref{elasteq}) could be 
decomposed as
\begin{equation}
\bmit u (\bmit x,t) = 
\sum_{n} A_{n}(t) \bmit\psi_{n}(\bmit x)
\label{decomposition}
\end{equation}
in which $\bmit\psi_{n}(\bmit x)$ are the eigenfunctions with 
eigenvalue $\omega_n$ of
\begin{equation}
\frac{\mu}{\rho} \ {\bmit \nabla}^2 \bmit \psi_{n}(\bmit x) + 
\frac{( \lambda + \mu)}{\rho}\ {\bmit \nabla} 
({\bmit \nabla} \cdot \bmit \psi_{n}(\bmit x)) =
- \omega ^2 _{n} \bmit \psi_{n}(\bmit x)  
\label{eigeneq}
\end{equation}
The eigenfunctions $\bmit\psi_{n}(\bmit x)$ are a complete set of 
orthonormal functions with the scalar product defined as
($^*$ means complex conjugation)
\begin{equation}
<\bmit g(r,\theta,\phi)|\bmit f(r,\theta,\phi)> = 
{1\over M}\int \bmit g^*\cdot\bmit f \rho d^3x
\label{scalarprod}
\end{equation}
where $M$ is the mass of the elastic body. Therefore
\begin{equation}
{1\over M}\int \bmit\psi^*_{n}\cdot\bmit\psi_{n'} \rho d^3x =
\delta_{nn'}
\label{orthonorm}
\end{equation}
Time dependent coefficients $A_n(t)$ satisfies the following 
forced harmonic oscillator equations
\begin{equation}
{\ddot A}_{n} + \omega ^2 _{n}\ A_{n} = R_n
\end{equation} 
where
\begin{equation}
R_{n}(t) = 
{1\over M}\int \bmit\psi^*_{n}(\bmit x) \cdot 
\bmit R(\bmit x,t) \rho d^3x
\label{gravacc}
\end{equation}

In this way a field equation has been transformed in a set of 
equations for the coefficients of a function series.
Now we can write the Lagrangian from which these equations are 
derived. This Lagrangian should be a real expression of the 
generalized coordinates $A_{n}(t)$ and $A_{n}^*(t)$. 
As far as kinetic energy of the elastic body (EB) is concerned, 
one has
\begin{equation}
T_{EB} = {1\over 2} \int d^3x \rho \dot{\bmit u}^2 = 
{1\over2} M \sum_{n} \dot A_{n} \dot A^*_{n}
\label{kineticeb}
\end{equation}
where last equality follows from~(\ref{orthonorm}).
Potential energy of both the elastic body and the external forces 
are immediately written as
\begin{eqnarray}
V_{EB} &=& {1\over 2} M\sum_{n} \omega^2_{n} A_{n} A^*_{n}
\label{potentialeb} \\
V_{ext} &=& - {1\over 2} M\sum_{n} 
\left [A_{n} R^*_{n} + A^*_{n} R_{n}\right ]
\label{potentialext}
\end{eqnarray}

\section{Lagrangian of the free sphere} 

If the elastic body is a sphere with radius $R$, then 
(see~\cite{ashdre75})
\begin{equation}
\bmit u (r,\theta,\phi,t) = 
\sum_{nml} A_{nml}(t) \bmit\psi_{nml}(r,\theta,\phi)
\label{sphdecomposition}
\end{equation}
where
\begin{eqnarray}
\bmit \psi_{nlm}(r,\theta,\phi) &=& 
[a_{nl}(r)\bmit n + b_{nl}(r)R \bmit{\nabla}] Y_{lm}(\theta,\phi)
\label{psisph}\\
a_{nlm} \left( r \right) & = & c_{nlm}\ R\  {d j_l \left(
q_{nl} r \right) \over d r} + d_{nlm} \ R\ 
{l(l+1) \over r} j_l \left(k_{nl} r \right) 
\label{radialspha} \\
b_{nlm} \left( r \right) & = &  c_{nlm} 
j_l \left(q_{nl} r \right) + d_{nlm} 
{d \over d r } \left( r j_l \left(k_{nl} r \right) \right) 
\label{radialsphb}
\end{eqnarray}
and $A_{nlm}(t) = (-1)^m A_{nl-m}^*(t)$.

Now we briefly sketch the procedure used by~\cite{wagpai77} to take
into account direction of the gravitational signal.
In the laboratory frame $O$, with origin in the centre of the sphere, 
the gravitational wave is coming by a given direction. Let us consider 
a reference frame $O'$ rotated with respect to $O$ in such a way that 
$z'$ axis coincides with the direction of propagation. All the primed
quantities refer to this last system. The radial component of the 
surface displacement in the primed frame is then given by 
(see~\cite[eq. (7)]{wagpai77})
\begin{equation}
\bmit n'\cdot\bmit u'(R,\theta',\phi') = 
\sum_{nlm'} a_{nl}(R) A'_{nlm'}(t) Y_{lm'}(\theta',\phi')
\label{normaldisplp}
\end{equation}
We are interested in the radial displacement in the laboratory frame.
Let us suppose that $O$ is rotated with respect to $O'$ by the hour 
angle $H$ (the angle between the projection of $z'$ axis on the plane
$xy$ and $x$ axis) and the declination $\delta$ (the angle between 
$z'$ axis and the plane $xy$, diminished by $2\,\pi$ if 
obtuse)~\cite{wagpai77}. In $O$ we have
\begin{equation}
\bmit n\cdot\bmit u(R,\theta,\phi) = 
\sum_{nlm} a_{nl}(R)
\left (\sum_{m'} A'_{nlm'} {\cal D}^{(l)}_{mm'}(H,\delta)\right )
Y_{lm}(\theta,\phi)
\label{normaldispl}
\end{equation}
where ${\cal D}^{(l)}_{mm'}(H,\delta)$ are the rotation matrix for 
spherical harmonics (see e.g.~\cite{edm57}).
Let us define
\begin{equation}
F_{nlm} = \sum_{m'} A'_{nlm'}{\cal D}^{(l)}_{mm'}(H,\delta)
\label{coefff}
\end{equation}
which are the coefficients of decomposition~(\ref{sphdecomposition}) 
in $O$. These coefficients generalizes in a natural way coefficients
$F_m$ as defined in~\cite[eq. (9)]{wagpai77}, for every 
order in indices $n$ and $l$. More precisely one has $F_m = F_{12m}$.
${\cal D}^{(l)}_{mm'}(H,\delta)$ are unitary matrices, therefore it is
possible to invert eq.~(\ref{coefff}) and find $A'_{nlm'}$ as linear 
combinations of $F_{nlm}$
\begin{equation}
A'_{nlm'} = \sum_{m} {\cal D}^{(l)\ *}_{mm'}(H,\delta) F_{nlm}
\label{apvsf}
\end{equation}

The kinetic energy of the sphere (S) as a function of the generalized
coordinates in the laboratory frame will be given by 
(see~(\ref{kineticeb}))
\begin{equation}
T_S = {1\over2} M \sum_{nlm'} \dot A'_{nlm'} \dot A'^*_{nlm'} =
{1\over2} M \sum_{nlm} \dot F_{nlm} \dot F^*_{nlm}
\label{kineticsph}
\end{equation}
As far as the elastic potential energy of the sphere and the potential 
energy of the external gravitational force are concerned one has 
\begin{eqnarray}
V_S &=& {1\over 2} M\sum_{nlm'} \omega^2_{nl} A'_{nlm'} A'^*_{nlm'} =
{1\over 2} M \sum_{nlm} \omega^2_{nl} F_{nlm} F^*_{nlm}
\label{potentialsph}\\
V_{ext} &=& - {1\over 2} M\sum_{nlm'} 
\left [A'_{nlm'} R^*_{nlm'} + A'^*_{nlm'} R_{nlm'}\right ] = 
- {1\over 2} M\sum_{nlm} \left [ G^*_{nlm} F_{nlm} + 
G_{nlm} F^*_{nlm}\right ]
\label{externalsph}
\end{eqnarray}
where we have set
\begin{equation}
G_{nlm} = \sum_{m'} {\cal D}^{(l)}_{mm'} R_{nlm'}
\label{gvsrp}
\end{equation}
which are the coefficients of the decomposition of the gravitational 
force in $O$.

\section{Coupling with transducers}
\label{coupling}
\subsection{General Case}

Let us consider a set of $N$ harmonic mechanical oscillators with mass 
$m_i$ and resonant frequency $\omega_i$, placed on the surface of the
sphere in $(R,\theta_i,\phi_i)$. 
Let $q_i(t)$ be the generalized coordinates which describe radial 
motion of each oscillator with respect to the surface of the sphere.
We make the assumption that $m_i$ are small with respect to $M$, the 
mass of the sphere. In this way we suppose that the presence of 
secondary oscillators does not modify the eigenfunctions of the sphere
(in the same way one suppose that the transducer of a resonant bar 
does not modify the eigenfunction of the cylinder).
Purpose of this section is to write that part of Lagrangian relative 
to the motion of the resonators and to their interaction with the 
surface of the sphere.

Concerning the potential energy of the transducers (t), the choice of 
generalized coordinates brings to the expression
\begin{equation}
V_t = \frac{1}{2} \sum_i m_i \omega_i^2 q_i^2
\label{potentialt}
\end{equation}
Moreover, if we call $y_i$ the coordinates of the i--th oscillator in 
the laboratory frame, then kinetic energy of transducers is written as
\begin{equation}
T_t = {1\over 2} \sum_i m_i \dot y_i^2
\label{kinetictlab}
\end{equation}
The relation between inertial and generalized coordinates is given by
\begin{equation}
q_i = y_i - \bmit n \cdot \bmit u(R,\theta_i,\phi_i,t) 
\stackrel{(\ref{normaldispl}), (\ref{apvsf})}{=} 
y_i - \sum_{nlm} a_{nl}(R) F_{nlm} Y_{nlm}(\theta_i,\phi_i)
\end{equation}
For convenience purposes, we define, for every $l$, the matrix 
$\bmit P^{(l)}$, whose components are
\begin{equation}
P^{(l)}_{mi} = Y^*_{lm}(\theta_i,\phi_i)  
\label{pmatrix}
\end{equation}
The connection between these matrices and the matrix $\bmit B$ of 
ref.~\cite{johmer93} is given by $\bmit B = \bmit P^{(2)*}$
With these positions the kinetic energy of the transducers is
\begin{eqnarray}
T_t & = & {1\over2} \sum_i m_i \bigg [
\dot q^2_i + \dot q_i \sum_{nlm} a_{nl}(R)
\left (P^{(l)\ *}_{mi} \dot F_{nlm} +
P^{(l)}_{mi} \dot F^{\ *}_{nlm} \right ) + \nonumber \\
& & \ \ \ \sum_{nlm} \sum_{n'l'm'} a_{nl}(R) a_{n'l'}(R) P^{(l)\ *}_{mi}
P^{(l')}_{m'i} \dot F_{nlm} \dot F^{\ *}_{n'l'm'} \bigg ]
\label{kinetict}
\end{eqnarray}

In conclusion the Lagrangian of an elastic sphere with $N$ mechanical 
oscillators on its surface, undergoing an external force is
\begin{equation}
L = T_S + T_t - V_S - V_t - V_{ext}  
\label{lagrnagiansph}
\end{equation}
where terms involved have been defined in~(\ref{kineticsph}),
(\ref{kinetict}), (\ref{potentialsph}), (\ref{potentialt}) 
and~(\ref{externalsph}) respectively.
Generalized coordinates are $F_{nlm}$, $F^{\ *}_{nlm}$ and $q_i$,
where $l=\{1,2,...\}$, $m=\{-l,-l+1,...,l\}$, and $i=\{1,2,..,N\}$.
Since $F_{nl-m} = (-1)^m F^{\ *}_{nlm}$ it is sufficient to find 
Lagrangian equations from $q_i$ and, say, $F^{\ *}_{lm}$.
Therefore, to simplify the problem, we can take suitable linear 
combinations of the coefficients $F_{nlm}$ in such a way to obtain real 
quantities. In particular, for every $l$ it exists a unitary operator 
$\bmit U^{(l)}$ which transforms irreducible tensors of rank $l$ in 
real vectors having $2l+1$ components. 
We will use lower case letters to indicate those real vectors
corresponding to irreducible tensors. One can therefore write
\begin{equation}
\bmit f_{nl} = \bmit U^{(l)} \bmit F_{nl}
\label{fvsF}
\end{equation}
where $\bmit f_{nl}$ is a vector whose components are $f_{nla}$
$(a \in \{1,2,\ldots,2l+1\})$, while 
$F_{nlm}$$(m \in \{-l,-l+1,\ldots,l\})$ are the components of
$\bmit F_{nl}$. From now on indices $a$, $b$, $c$ run from
$1$ to $2l + 1$. In this way, for instance, components of 
$\bmit U^{(l)}$ will be given by $U^{(l)}_{am}$. A possible choice for 
such unitary operators is the following
\begin{eqnarray}
U^{(l)}_{1\, m} & = & \delta_{m\, 0} \nonumber \\
U^{(l)}_{2 m'\, m} & = & - \frac{i}{\sqrt{2}}\, \left (
\delta_{m'\, m} - (-1)^m \delta_{m'\, -m} \right ) 
\label{unitaryu}\\
U^{(l)}_{2 m'+1\, m} & = & \frac{1}{\sqrt{2}}\, \left (
\delta_{m'\, m} + (-1)^m \delta_{m'\, -m} \right ) \nonumber
\end{eqnarray}
where $m' \in \{1,2,\ldots,l\}$. 
Performing these unitary transformation Lagrangian is obtained as a 
function of the real and independent generalized coordinates $f_{nla}$ 
and $q_i$. With this new set of variables, kinetic and elastic 
potential energy of the sphere, kinetic and potential energy of the 
transducers and potential energy of the external gravitational force 
acting on the sphere become respectively
\begin{eqnarray}
T_S &=& \frac{1}{2} M \sum_{nla} \dot f_{nla}^2
\label{kineticsphreal} \\
V_S &=& \frac{1}{2} M \sum_{nla} \omega_{nl}^2 f_{nla}^2
\label{potentialsphreal} \\
T_t & = & {1\over2} \sum_i m_i \bigg [
\dot q^2_i + 2 \dot q_i \sum_{nla} a_{nl}(R)
p^{(l)}_{ai} \dot f_{nla} + \nonumber \\
& & \ \ \ + \sum_{nla} \sum_{n'l'a'} a_{nl}(R) a_{n'l'}(R) p^{(l)}_{ai}
p^{(l')}_{a'i} \dot f_{nla} \dot f_{n'l'a'} \bigg ]
\label{kinetictreal} \\
V_t &=& \frac{1}{2} \sum_i m_i \omega_i^2 q_i^2
\label{potentialtreal} \\
V_{ext} &=& - M \sum_{nla} g_{nla} f_{nla}.
\label{potentialextreal}
\end{eqnarray}
where the matrix $\bmit p^{(l)} = (p^{(l)}_{ai})$ is given by
\begin{equation}
p^{(l)}_{ai} = \sum_m U^{(l)}_{am} P^{(l)}_{mi}.
\label{plai}
\end{equation}
Obviously the potential energy of the transducers is left unchanged.
Lagrangian equations
\begin{eqnarray*}
\frac{d\ }{dt} \left ( \frac{\partial L}{\partial \dot f_{nla}}
\right ) - \frac{\partial L}{\partial f_{nla}} & = & 0 \\
\frac{d\ }{dt} \left ( \frac{\partial L}{\partial \dot q_{i}}
\right ) - \frac{\partial L}{\partial q_{i}} & = & 0 
\end{eqnarray*}
give therefore the following equations of motion for the 
modes of a forced elastic sphere in interaction with $N$ harmonic
mechanical oscillators 
\begin{eqnarray}
M \ddot f_{nla} + M \omega_{nl}^2 f_{nla} - 
\sum_i a_{nl}(R) m_i \omega_i^2 p^{(l)}_{ai} q_i & = & M g_{nla}
\nonumber \\
\label{equationsnlm} \\
m_i \ddot q_i + m_i \sum_{nla} a_{nl}(r) p^{(l)}_{ai} \ddot f_{nla} +
m_i \omega_i^2  q_i & = & 0 \nonumber
\end{eqnarray}

\subsection{Quadrupolar Interaction}

In which follows we make the simplifying assumptions that secondary
oscillators are identical, their common frequency is equal to 
$\omega_{12}$ (the frequency of the mode $n=1$,\, $l=2$ of the 
sphere), their mass is $m$ and the external force acting on the sphere 
has only $l=2$ components. Moreover, for high quality factors 
($\tau_{nlm}$ is the relaxation time of the $nlm$ mode)
\[
| \omega_{12} - \omega_{nl} | >> {1\over\tau}\qquad\qquad
\tau = min \{\tau_{12m},\tau_{nlm'}\};\qquad\qquad
\begin{array}{c}
n \neq 1 \\
l \neq 2
\end{array}
\]
therefore if the motion of the transducers is studied in the frequency 
domain near the frequency $\omega_{12}$, only modes with $n=1$ and 
$l=2$ must be taken into account. To simplify notations we set
\[
p^{(2)}_{ai}\equiv p_{ai}
\qquad
f_{12a} \equiv f_{a}
\qquad
g_{12a} \equiv g_{a}
\qquad
\omega_{12} = \omega_0
\qquad
a_{12}(R) \equiv \alpha
\] 
With these assumptions we have
\begin{eqnarray}
T_S &=& {1\over2} M \sum_{a} \dot f_{a}^2
\label{kineticsph2} \\
V_S &=& {1\over 2} M \omega_0^2 \sum_{a} f_{a}^2
\label{potentialsph2} \\
T_t &=& {m\over2} \sum_i \bigg [
\dot q^2_i + 2 \dot q_i \sum_{a} \alpha
p_{ai} \dot f_{a} +
\sum_{aa'} \alpha^2  p_{ai}
p_{a'i} \dot f_{a} \dot f_{a'} \bigg ]
\label{kinetict2} \\
V_t &=& {m\omega^2\over 2} \sum_i q_i^2
\label{potentialt2} \\
V_{ext} &=& - {1\over 2} M \sum_{a} g_a f_a
\label{potentialext2}
\end{eqnarray}
by means of which the following equations of motions are obtained
($^t$ means transpose)
\begin{equation}
\left (
\matrix {M{\cal I}_5 & 0
\cr
m\alpha\bmit p^t & m \bmit{\cal I}_N
\cr}
\right ) 
\left (
\matrix {\ddot{\bmit f}
\cr
\ddot{\bmit q}
\cr}
\right )
+
\left (
\matrix {M\omega_0^2 \bmit{\cal I}_5 & -m\omega_0^2\alpha\bmit p
\cr
0 & m\omega_0^2 \bmit{\cal I}_N
\cr}
\right ) 
\left (
\matrix {\bmit f
\cr
\bmit q
\cr}
\right )
=
\left (
\matrix {\bmit{\cal I}_5 & 0
\cr
0 & \bmit{\cal I}_N
\cr}
\right ) 
\left (
\matrix {M \bmit g
\cr
0
\cr}
\right )
\label{equations12m}
\end{equation}
where $\bmit{\cal I}_n$ is the identity matrix in n--dimensional 
space.

If there are forces $r^{(ns)}_i$ between oscillators and the surface 
of the sphere (for instance thermal noise forces), then a new 
potential energy term must be added to the Lagrangian
\begin{equation}
W_{ns} = - \sum_i q_i r^{(ns)}_i
\label{noisepotential}
\end{equation}
The equations of motion then become (cfr~\cite[eq. (25)]{merjoh95})
\begin{equation}
\left (
\matrix {M \bmit{\cal I}_5 & 0
\cr
m\alpha\bmit p^t & m \bmit{\cal I}_N
\cr}
\right ) 
\left (
\matrix {\ddot{\bmit f}
\cr
\ddot{\bmit q}
\cr}
\right )
+
\left (
\matrix {M\omega_0^2 \bmit{\cal I}_5 & -m\omega_0^2\alpha\bmit p
\cr
0 & m\omega_0^2 \bmit{\cal I}_N
\cr}
\right ) 
\left (
\matrix {\bmit f
\cr
\bmit q
\cr}
\right )
=
\left (
\matrix {\bmit{\cal I}_5 & -\alpha \bmit p
\cr
0 & \bmit{\cal I}_N
\cr}
\right ) 
\left (
\matrix {M \bmit g
\cr
\bmit r^{(ns)}
\cr}
\right )
\label{equations12mns}
\end{equation}
The coefficient $\alpha = a_{12}(R)$ can be calculated 
from~(\ref{radialspha}) solving eigenfunction 
equations~(\ref{eigeneq}). One find that $\alpha = -2.886$.

Now the system of $5+N$ coupled oscillators must be decoupled thus
finding the eigenfrequencies. It is important to check if the 
frequency spread due to the decoupling is consistent with the 
assumption that only modes with $n=1$ and $l=2$ have been taken into 
account. Calling $\tilde\omega_a$ these eigenfrequencies and
$\tilde\tau_a$ the relaxation times of the eigenmodes, 
then the following consistency relation should hold
\[
| \tilde\omega_a - \omega_{nl} | >> {1\over\tau}\qquad\qquad
\tau = min \{\tilde\tau_a,\tau_{nlm}\}\qquad\qquad
\begin{array}{c}
n \neq 1 \\
l \neq 2
\end{array}
\]
As it will be seen, the above condition is fulfilled for high quality 
factors and $m_i/M<<1$.
To perform such a program a particular arrangement of transducers must 
be chosen. It is what we are going to do in the next section.

\section{Dodecahedron arrangement}

Now, the problem of the decoupling of the system made by the 
quadrupolar modes of the sphere interacting with N harmonic 
oscillators is to be solved. In order to do so the number and exact 
position of the mechanical oscillators on the surface must be known.
This choice is of fundamental importance to keep isotropy in the 
sensitivity. Until now, two arrangements in particular are studied
in view of an experimental realization. 
The first one~\cite{johmer93,merjoh95} is obtained placing six radial 
resonators along the directions of a dodecahedron (this arrangement 
was yet investigated~\cite{ort92a,ort92b,ort93} in the framework of 
the study of a network of resonant gravitational antennae).
In the second one~\cite{zoumic95} 5 transducers (four of which 
tangential) are placed in such a way that they could be excited by 
only one mode of the sphere; in this way decoupling is realized at 
once. In this paper we restrict ourselves to the first case.

Let us consider therefore the dodecahedron arrangement. Five transducers 
are placed on the same parallel of the sphere. Their common azimuth 
angle is $\theta =$ $\arcsin(2/\sqrt{5})$; their polar angles are 
given by $\phi_k = 2k \pi/5$ ($k \in \{1,\ldots,5\}$). Last transducer 
is placed at a pole. In this case matrix $\bmit p$ become
\begin{equation}
\bmit{p} = \sqrt{\frac{3}{5 \pi}}\, \left (
\begin{array}{cccccc}
1 & c_4 & c_2 & c_2 & c_4 & 0 \\
0 & -s_4 & s_2 & -s_2 & s_4 & 0 \\
-1 & -c_2 & -c_4 & -c_4 & -c_2 & 0 \\
0 & s_2 & s_4 & -s_4 & -s_2 & 0 \\
-a_1 & -a_1 & -a_1 & -a_1 & -a_1 & 5 a_1
\end{array} \right ) \qquad
\begin{array}{ccc}
c_2 = \cos{\left ( \frac{2 \pi}{5} \right )} 
& \qquad &
c_4 = \cos{\left (\frac{4 \pi}{5} \right)} \\
s_2 = \sin{\left (\frac{2 \pi}{5}\right )} 
& \qquad & 
s_4 = \sin{\left (\frac{4 \pi}{5}\right )} \\
a_1 = \frac{1}{2 \sqrt{3}} & \qquad & \qquad
\end{array}
\label{pp}
\end{equation}
Such a matrix has the following important properties 
(cfr~\cite{johmer93,merjoh95})
\begin{equation}
\bmit p \bmit p^t = \frac{3}{2 \pi} \bmit{\cal I}_5;
\qquad\qquad
\sum_{i} p_{ai} = 0
\label{pprop}
\end{equation}

Decoupling system~(\ref{equations12mns}) means solving an eigenvalue 
problem. The eigenfunctions are the so called normal coordinates, 
while the eigenfrequencies are called frequencies of free vibration
(see~\cite{gold80}). In the case of this arrangement, normal 
coordinates are divided in three set: two quintuplet with 
eigenfrequencies $\omega_+ = \omega_0\,\lambda_+$ and 
$\omega_- = \omega_0\,\lambda_-$ and a singlet with 
frequency $\omega_0$ where
\begin{equation}
\lambda_\pm = 
1 \pm \frac{1}{2} \sqrt{\frac{3 \alpha^2}{2 \pi} \mu} 
\qquad\qquad \mu = \frac{m}{M}
\label{lambdamp}
\end{equation}
The normal coordinates $\zeta^p$ ($p=1,\dots,11$) are therefore 
naturally divided in three groups: $\bmit \zeta_- = (\zeta_p)$
$p \in \{1,\ldots,5\}$, $\zeta = \zeta_6$ and 
$\bmit \zeta_+ = (\zeta_i)$ $p \in \{7,\ldots,11\}$.
The equations of forced motion are (detailed calculations are found in 
appendix~\ref{decoupling})
\begin{eqnarray}
\ddot{\bmit\zeta}_- +\lambda_-^2 \omega_0^2 \bmit\zeta_- &=&
\sqrt{\frac{M}{2}} \lambda_- \bmit g \nonumber \\
\ddot \zeta + \omega_0^2 \zeta &=& 0 \label{equationsnm} \\
\ddot{\bmit\zeta}_+ +\lambda_+^2 \omega_0^2 \bmit\zeta_+ &=&
\sqrt{\frac{M}{2}} \lambda_+ \bmit g \nonumber
\end{eqnarray}
Therefore if $\bmit g$ is given, one can 
find the motion of coefficients $f_a$ and transducers $q_i$. In 
particular because of the form of matrix $\bmit A$ (see 
eq.~(\ref{aaa})) we have
\begin{eqnarray}
\bmit f &=& \sqrt{\frac{1}{2 M}}\,
\left (
\lambda_- \bmit \zeta_- + \lambda_+ \bmit \zeta_+
\right )
\nonumber \\
\label{conv} \\
\bmit q &=& \sqrt{\frac{\pi}{3 m}}\, \bmit p^t \, 
\left (
\lambda_- \bmit \zeta_- - \lambda_+ \bmit \zeta_+
\right ) + \frac{\zeta}{\sqrt{6 m }} \bmit 1_{(6 \times 1)}
\nonumber
\end{eqnarray}

\section{Decomposition of the gravitational wave force}
\label{gravity}

A gravitational wave, in the framework of a general metric theory of 
gravity, is described as a combination of six matrices. A possible 
choice is the following (see also~\cite{lobo95}):
\begin{eqnarray}
\bmit E^{(S)} &=& \frac{2}{\sqrt{15}} \left (
\begin{array}{ccc}
1 & 0 & 0 \\
0 & 1 & 0 \\
0 & 0 & 1 
\end{array} \right) \label{scalar} \\
\bmit E^{(0)} = \frac{1}{\sqrt{3}} \left (
\begin{array}{ccc}
-1 & 0 & 0 \\
0 & -1 & 0 \\
0 & 0 & 2 
\end{array} \right ) \qquad
\bmit E^{(\pm 1)} &=& \left (
\begin{array}{ccc}
0 & 0 & 1 \\
0 & 0 & \pm i \\
1 & \pm i & 0
\end{array} \right ) \qquad
\bmit E^{(\pm 2)} = \left (
\begin{array}{ccc}
1 & \pm i & 0 \\
\pm i & -1 & 0 \\
0 & 0 & 0 
\end{array} \right ) 
\label{quadrupolar}
\end{eqnarray}
Matrix~(\ref{scalar}) accounts for the scalar part of the signal, while 
the others describe the quadrupolar components of the wave. They are 
relative to $m=0$, $m=\pm 1$ and $m=\pm 2$ respectively.
One may argue that, once the direction of propagation is known, a 
sphere with dodecahedron arrangement could detect, at least in 
principle, all six components. However, the scalar mode of the sphere 
has a principal resonant frequency $\omega_{10}$, which is  a little 
more than twice the quadrupolar one $\omega_{12}$ 
(e.g.~\cite{lobo95}). Current experimental schemes needs that, 
in order to measure the scalar mode of the sphere, this must be 
coupled with a secondary oscillator having $\omega_{10}$ as a resonant 
frequency. Therefore, being instead $\omega_{12}$ the resonant 
frequency of each transducer, scalar part of the signal cannot be 
detected. Besides, propagation direction is in general unknown. One is 
led to the conclusion that with a dodecahedron 
arrangement it is possible to get the direction of propagation and 
three components of the five quadrupolar possible ones. In the framework 
of general relativity, a gravitational wave could only have two possible 
states of polarization, the $\bmit E^{(\pm 2)}$ ones. 
Therefore one can think that a sphere could be used to test general 
relativity. However, the discovery of PSR1913+16~\cite{hultay75} and 
20 years of subsequent observations have put strong experimental 
limitations on other theory of gravitation different from general 
relativity. This last one passed all these new experimental tests with 
complete success~\cite{psr} showing that ``the correct theory of 
gravity must make predictions that are asymptotically close to those 
of general relativity over a vast range of classical 
circumstances''~\cite{tay94}. Therefore another possibility is to 
consider general relativity as the correct metric theory of gravity 
and then to use the degree of  freedom left as a veto against spurious 
signals. This is the strategy of the present work. In which follows we 
consider a gravitational wave in $O'$ having the form
\begin{equation}
\bmit E'(t) = E_R(t) \bmit E^{(+2)} + 
E_L(t) \bmit E^{(-2)} + 
E_l(t) \bmit E^{(0)}
\label{gravwave}
\end{equation}
where $E_L = E_R^*$ and the connection with the usual polarization 
amplitudes is $E_+(t) = E_R(t) + E_L(t)$ and 
$E_\times (t) = i \left (E_R(t) - E_L(t) \right )$, while $E_l$ is the 
eventual longitudinal polarization.

With this position it is possible to find, for every $n$, the vector 
$\bmit g_n$ (whose components are $g_{n2a}$) which enters in the 
equations of motions of the normal modes~(\ref{equationsnm}) as a 
function of $\bmit E'(t)$ by means of eqs.~(\ref{fvsF}), 
(\ref{gvsrp}), (\ref{gravacc}) and (\ref{gravityforce}). 
Detailed calculations are found in appendix~\ref{force}. Here we 
give only the final result
\begin{eqnarray}
(g_n)_5(t) &=& \frac{\gamma_n}{\sqrt{2}} \left [
\frac{1}{4} \left (3-\cos 2 \delta \right ) \cos 2 H\ E_+ -
\sin \delta \sin 2 H\ E_\times + \right. \nonumber \\
& &\qquad\qquad\qquad\qquad\qquad\qquad\qquad\qquad \left.
+\ \frac{\sqrt{3}}{4} \left (1+\cos 2 \delta \right ) \cos 2 H\ E_l
\right ] \nonumber \\
(g_n)_4(t) &=& \frac{\gamma_n}{\sqrt{2}} \left [
\frac{1}{4} \left (3-\cos 2 \delta \right ) \sin 2 H\ E_+ +
\sin \delta \cos 2 H\ E_\times + \right. \nonumber \\
& &\qquad\qquad\qquad\qquad\qquad\qquad\qquad\qquad \left.
+\ \frac{\sqrt{3}}{4} \left (1+\cos 2 \delta \right ) \sin 2 H\ E_l
\right ] \nonumber \\
(g_n)_3(t) &=& \frac{\gamma_n}{\sqrt{2}} \left (
\frac{1}{2} \sin 2 \delta \cos H\ E_+ -
\cos \delta \sin H E_\times -
\frac{\sqrt{3}}{2} \sin 2 \delta \cos H\ E_l \right)
\nonumber \\
(g_n)_2(t) &=& \frac{\gamma_n}{\sqrt{2}} \left (
\frac{1}{2} \sin 2 \delta \sin H\ E_+ +
\cos \delta \cos H E_\times -
\frac{\sqrt{3}}{2} \sin 2 \delta \sin H\ E_l \right)
\nonumber \\
(g_n)_1(t) &=& \frac{\gamma_n}{\sqrt{2}} \left [
\frac{\sqrt{3}}{4} \left (1+\cos 2 \delta \right ) E_+ +
\frac{1}{4} \left (1-3\cos 2 \delta \right ) E_l
\right ] \label{vectorg} 
\end{eqnarray}
where $\gamma_n$ is defined in appendix~\ref{force}.

It has been therefore explicitly found the motion of mechanical
transducers on the surface of the sphere as a function of direction 
and polarization amplitudes of a signal (by means of~(\ref{conv}), 
(\ref{equationsnm}) and (\ref{vectorg})), solving the direct problem 
of motion.

\section{Inverse problem}

Inverse problem lies in the determination of the incoming signal 
characteristics 
for a given response of the detection apparatus. In this case what is 
measured is the $6$ component vector $\bmit q(t)$, that is the motion 
of transducers with respect to the surface of the sphere. From this 
vector one should obtain $\bmit \zeta(t)$, describing the motion of 
normal modes, and $\bmit g$, the generalized force for the normal 
coordinates. Finally, inverting system~(\ref{vectorg}) one should 
get those quantities which describe the signal, that is direction, 
polarizations $E_+$ and $E_\times$ and an eventual polarization with
$m=\pm 1$ or $m=0$. In which follows we consider only the possibility 
that the signal could have a longitudinal polarization $E_l$ 
(see~(\ref{gravwave})).

In Fourier space 
($f(\omega)=$ $(2\pi)^{-1/2}\int_{-\infty}^{+\infty} 
f(t) \exp(-i\omega t) \ dt$) using properties~(\ref{pprop}) of matrix 
$\bmit p$ and eq.~(\ref{equationsnm}) one gets
\begin{equation}
\bmit g(\omega) = - \frac{2 \pi}{3 \alpha}\ 
\frac{\left (\omega^2_0 - \omega^2 \right )^2}{\omega^2}\ 
\bmit p^t \bmit q(\omega)
\label{decon}
\end{equation} 
This last equation implies, together with~(\ref{pprop}) 
\begin{equation}
\bmit q^t(\omega) \bmit q(\omega) \propto 
\bmit g^t(\omega) \bmit g(\omega) = 
\frac{\gamma_1^2}{2} \left (E_+^2 + E_\times^2 + E_l^2 \right )
\label{isotropy}
\end{equation}
where last equality holds taking into account eqs.~(\ref{fvsF}),
(\ref{gvsrp}) and (\ref{Rdz}).
Therefore the sum of the squares of the detector responses are 
independent on the direction. This means that this transducer 
arrangement gives isotropic sensitivity.
The result could be also obtained calculating directly the sum of the 
squares of the components of $\bmit g$ in~(\ref{vectorg}).

Inverse problem is then solved when system~(\ref{vectorg}) is 
inverted. To this aim it is useful to define an amplitude vector 
$\bmit a$
\begin{equation}
\bmit a = \frac{\sqrt{2}}{\gamma_1} \bmit g,
\qquad\qquad
\bmit a^2 = E_+^2 + E_\times^2 + E_l^2,
\label{vectora}
\end{equation}
whose length is the sum of the squares of the three polarization 
amplitudes of the signal.
Performing this program (results are found in 
appendix~\ref{inversion}) one obtains a six-th degree algebraic 
equation for $x = \tan{H/2}$
with coefficients depending on $\bmit a$. Then a relation for 
$\tan{\delta}$ as a function of $H$ and $\bmit{a}$; finally three 
relations for $E_+$, $E_\times$, $E_l$ as functions of $H$, 
$\delta$ and $\bmit{a}$.

As it will be seen better in the following, since for a gravitational 
signal $E_l$ must vanish, only two of the six solutions for $H$ can be 
accepted. In fact longitudinal components will be zero for only two 
values of $H$. These acceptable solutions differ only because they 
describe signals coming along opposite directions. In fact the 
difference between their hour angles is $\pi$. This property 
holds also for the discarded solutions: they are in pairs
differing only for the opposite direction of propagation.

Solution of the inverse problem lies therefore in the solution of the 
equation for $H$. This could be done numerically once vector $\bmit a$ 
is known. To this purpose a code has been developed which solves the 
direct and the inverse problem. The obtained calculation precision 
on the reconstruction of the signal is very much higher than the 
errors introduced by thermal noise which we are going to consider in 
next section. 

\section{Thermal noise}

So far the theory dealt with ideal detectors, that is to say 
noiseless ones. However, as it is well known, real detectors are not 
at rest before the interaction with a signal but, because of the heat 
bath in which they are thermodynamically in equilibrium, they have a 
Brownian motion which give them a certain amount of energy.
For this reason a detector (a part from the quantistic aspects of the 
problem, aspects which come out indeed when the temperatures are lower 
than those actually involved at the present) could only reveal signals 
yielding an amount of energy greater or at least comparable to the 
thermic one. 
Therefore, for any given temperature, there will be a lowest signal 
detectable.
In this section we analyze how the behaviour of the system described 
until now is to be modified by the presence of thermal noise.

To this aim, the problems that must be solved are at least two. The 
first one is to understand which is the noise that one could expect in 
the transducers on the surface of the sphere at a certain temperature.
The second one concerns the knowledge of the decreasing in the 
precision with which not only the signal could be deconvolved (by 
means of the determination of the arrival direction and polarization 
amplitudes) but also with which could be used the veto on the 
longitudinal part of the signal.
In other words a confidence level must be decided for which a certain 
value of $E_l$ (which could not be exactly zero) is to be considered 
equal to zero or not.

Let us begin to consider the first problem answering to the question: 
which noise is to be expected on the transducers at a given 
temperature? As a first estimate to be compared with the result of a 
more accurate calculation we may suppose, if the sphere mass is very 
much greater than that of the oscillators, that the Brownian noise 
cause a motion whose amplitude with respect the surface of the sphere  
is of the order of
\begin{equation}
q^{(0)}_i = \sqrt{\frac{2\,k_B\,T}{m\,\omega_0^2}}
\label{qzero}
\end{equation}
This result shall be valid as an order of magnitude, 
allowing settlements which, in general, depend upon the position of 
the oscillators on the surface of the sphere. We will see that, in the
dodecahedron configuration this result is indeed correct being the 
same for every transducer.

Strictly speaking, the equipartition energy theorem, that we used for 
giving the order of magnitude of the amplitudes $q^{(0)}_i$, is valid, 
in classical mechanics, only for every independent quadratic term in 
the energy and for every one of these the square--mean value is
$(1/2)\,k_B\,T$. Therefore, when we consider interacting oscillators 
we can apply it just to the normal coordinates.
It must be observed, however, that the theory developed until now is 
only an approximation: in fact together with the lagrangian of the 
elastic body in interaction with the secondary oscillators we have not 
consider the dissipation function~\cite[\S 1--5]{gold80}. As it is well 
known it is a quadratic function of the velocities
${\cal F} = (1/2) {\cal F}_{ij} \dot \eta^i \dot \eta^j$ and 
$2 {\cal F}$ is the rate of energy dissipation due to 
friction~\cite[\S 2.6]{gold80}. The motion equations 
become~\cite[\S 6.5]{gold80}
\[
T_{ij} \ddot \eta^j + {\cal F}_{ij} \dot \eta^j + V_{ij} \eta^j = 0
\]
It is not in general possible to find normal modes for any arbitrary 
dissipation function. In other words it is not possible to decompose
the system in non--interacting modes diagonalizing simultaneously the 
three matrices $T_{ij}$, $V_{ij}$ and ${\cal F}_{ij}$. 
Therefore the equipartition theorem could not be applied even to the 
normal coordinates. If we make the assumption of high quality factor 
we can suppose that non--diagonal quadratic terms are very small with 
respect to the others so we can safely neglect them.
In the framework of this approximation we can therefore apply the 
equipartition theorem to the normal coordinates, whose Brownian motion 
will then be given by (see~(\ref{equationsnm}))
\begin{eqnarray}
\bmit\zeta_- &=& \frac{\sqrt{2\,k\,T}}{\lambda_-\,\omega_0} 
\bmit n_-(t) \nonumber \\
\zeta &=& \frac{\sqrt{2\,k\,T}}{\omega_0} n(t) \label{nsmn} \\
\bmit\zeta_+ &=& \frac{\sqrt{2\,k\,T}}{\lambda_+\,\omega_0} 
\bmit n_+(t) \nonumber
\end{eqnarray}
where $\bmit n_-(t)$, $n(t)$ and $\bmit n_+(t)$ are eleven 
independent adimensional random functions (where ``random'' means that
their average values vanish) whose root--mean--square deviations are 
equal to one.
The relative coordinates $\bmit q(t)$ are given as a function of the 
normal ones $\bmit \zeta(t)$ by the equation~(\ref{conv}) which now 
could be written as
\begin{equation}
\bmit q = \frac{\sqrt{2\,k\,T}}{\omega_0} \left [
\sqrt{\frac{\pi}{3 m}}\, \bmit p^t \, 
\left (n_-(t) - n_+(t) \right ) +
\frac{n(t)}{\sqrt{6 m }} \bmit 1_{(6 \times 1)} \right ]
\end{equation}
Let us now define the following five functions
\begin{equation}
\bmit v(t) = \frac{1}{\sqrt{2}} \left (n_-(t) - n_+(t) \right )
\end{equation}
For these functions (from now on $<f>$ means the average of a given 
quantity $f(t)$) one obviously has that $<\bmit v>=0$ (that is to say 
they are random functions) and $<v_a^2> = 1$. We can therefore write
\begin{equation}
\bmit q(t) = \frac{1}{\omega_0} \sqrt{\frac{2\,k\,T}{3\,m}} \left [
\sqrt{2\,\pi} \bmit p^t\,\bmit v(t) + \frac{1}{\sqrt{2}}\, n(t)\,
\bmit 1_{(6 \times 1)} \right ]
\end{equation}
Let now be $\bmit w(t)$ six functions defined as
\begin{equation}
\bmit w(t) = \sqrt{\frac{2}{5}} \sqrt{2 \pi} \bmit p^t
\bmit v(t)
\end{equation}
As it is easily seen these are random functions
and we have also that $<w_i^2> = 1$. This follows directly from the 
fact that $(\bmit p^t\,\bmit p)_{ii} = 5/(4 \pi)$. 
This property of the matrix $\bmit p$ is of fundamental importance 
in order that all the transducers have the same noise amplitude.
Introducing these new functions we obtain
\begin{equation}
\bmit q(t) = \frac{1}{\omega_0} \sqrt{\frac{2\,k\,T}{3\,m}} \left [
\sqrt{\frac{5}{2}} \bmit w(t) + \sqrt{\frac{1}{2}}\, n(t)\,
\bmit 1_{(6 \times 1)} \right ]
\end{equation}
putting now
\begin{equation}
\bmit z(t) = \sqrt{\frac{1}{3}} \left (
\sqrt{\frac{5}{2}} \bmit w(t) + \sqrt{\frac{1}{2}}\, n(t)\,
\bmit 1_{(6 \times 1)} \right )
\end{equation}
where once again $<\bmit z>=0$ and $<z_i^2>=1$, one gets
\begin{equation}
\bmit q(t) = \frac{1}{\omega_0} \sqrt{\frac{2\,k\,T}{m}} \bmit z(t)
\end{equation}
We have therefore shown that the fluctuations due to the Brownian 
motion give rise to a random motion of the coordinates $\bmit q(t)$ 
with root--mean--value just given by~(\ref{qzero}).
It is to be remarked that this result depends on the particular 
arrangement of the secondary oscillators on the surface of the sphere. 
A priori one would in fact expect a position dependent result.

We have therefore given the answer to the first question put by the 
presence of noise. The next section is devoted to the solution of the 
second problem envisaged here: the reconstruction precision.

\section{Numerical solution}

In this section we are interested in the answer to the problem of 
knowing the decrease in the signal reconstruction precision when 
brownian motion is to take into account. It is to be remarked that 
this precision decrease concerns not only the arrival direction and 
the transverse polarization amplitudes, but also the value of the 
longitudinal component which, in case of gravitational event, must be 
zero but, because of noise, could not be ``mathematically'' so.
Therefore we have to precise the criterion with which this signal 
component is to be considered ``physically'' vanishing.
Because of the complexity of the system we can not simply apply a kind 
of analysis based on the error propagation: the best solution 
is to perform numerical simulations.

The strategy is to write a code which, once the signal 
characteristics are given, calculates the components of the vector
$\bmit a$ (that is to say the response of the device), gives them a 
gaussian noise with null mean value and unitary root--mean-square 
deviation and finally solves the inverse problem.
When this is done six solutions are found which represents a possible 
signal. The four solutions which have the greater longitudinal 
polarization are discarded. With an iterative procedure over $n$ 
attempts ($n>>1$), one could extract two kinds of information (for 
every value of the signal--to--noise ratio): the first one is about 
the error on the reconstruction of the initial signal; 
the second one is about the 
typical value of the ratio between the longitudinal component $E_l$ 
and the amplitude $E = (E_+^2 + E_\times^2)^{(1/2)}$ which one could 
expect from a real measure in the case of gravitational (that is to 
say transverse) signal.

We have reached the conclusion that the signal could be reconstructed 
when the signal--to--noise ratio ($SNR$) is equal or greater than $10$.
If this is the case the relative error on the amplitude $E$ and the 
absolute one on the $\delta$ angle is of the order of $1/SNR$.
A great lack of precision appears on the $H$ angle and on the 
polarization angle $\psi=\arctan{E_\times/E_+}$ (that is to say on the 
knowledge of the 
two polarization amplitude) when $\delta$ assumes values closer to 
$\pm \pi/2$, as it was anyway well-founded to be expected.
As far as the ratio $E_l/E$ is concerned, it is distributed around $0$ 
with a typical spread which is of the order of the reciprocal of 
$SNR$.
From this analysis one could conclude that, if the response of the 
detector lead to a reconstruction of the signal in which the six 
solutions are such that the ratio $E_l/E$ is significantly greater 
than $1/SNR$, then this signal is certainly to be considered a spurious 
event and rejected as a possible gravitational event.

In tables below we report some of the results and graphics 
concerning two simulations over 1000 attempts. 
In the first case the signal was supposed to come from a source placed 
at hour angle $H=1.0\,rad$ and declination $\delta=0.7\,rad$. The 
amplitude $E$ of the wave is $10$ (in units of the SNR), with the two 
polarization amplitudes given by $E_+ = 8$ and $E_\times = 6$.
In the second case the source is placed at hour angle $H=1.0\,rad$ and 
declination $\delta= 1.3\,rad$. The signal amplitude is still given by
$E = 10$, with $E_+ = 6$ and $E_\times = 8$.
The results of the simulations are given in table~\ref{tab1}.
The quantity $\Delta \Omega$ is the error on the solid angle; one has 
$\Delta \Omega = \Delta H\,\Delta\delta\,\cos{\delta}$.
In Figs.~\ref{fig1} and~\ref{fig2} are shown the scattered plots of 
reconstruction position in the two cases.

An analysis of the results shows immediately that the precision on $H$ 
decrease when the source become nearer to the poles. However this have 
not substantially repercussions on the solid angle precision.
Joined directly with the ``spreading'' of $H$ is the uncertainty on 
the polarization amplitudes which increases when the signal direction 
approaches a polar one.

A consequence of this analysis is that, once a reference frame $O$ is 
chosen, the isotropy of the sphere with 
thermal noise concerns direction, total amplitude $E$ and amplitude of 
the longitudinal component $E_l$ of the signal.
As far as the possibility of reconstruction of the polarization 
amplitudes $E_+$ and $E_\times$, it is strongly dependent on the 
declination angle. That is, sphere isotropy on the 
reconstruction of polarization angle $\psi=\arctan(E_\times/E_+)$ is 
broken by thermal noise near polar singularity.

\section{conclusions}

An elastic sphere could represent, in a no longer far future, an 
important device in the context of gravitational astronomy. This 
lies in its peculiarity of total reconstruction of the signal with 
isotropic sensitivity and in the possibility of testing the 
transversality of the Riemann tensor. This last fact is very important 
because it allows to exert a veto against spurious signals.
We think that the operation usefulness of a gravitational antenna does 
not depend only on its sensitivity or capability of signal 
reconstruction, but also on the confidence level with which an observed 
event could be associated to a gravitational wave 
(see~\cite{ort92a,ort92b,ort93}). The Lagrangian approach we have used, 
has been developed in the context of this strategy in order to write 
the correct equations of the coupled oscillators using the simpler 
formalism. In fact we think that a serious study should be done in 
order to find other configurations of the secondary oscillators for 
which also an eventual scalar amplitude of the signal could be measured 
and used as a veto. A first approach in this direction has been 
attempted with a different method by~\cite{lobser96}. 

The analysis of the simulated output in the presence of thermal noise 
has shown that an amplitude signal--to--noise ratio is needed of almost 
$10$, in order to exploit all the possibilities of a resonant sphere
(cfr~\cite{ort93}).

\acknowledgments

The author wish to thank P. Fortini and E. Coccia for useful 
discussions and J. A. Lobo for his encouragement.

\appendix

\section{Decoupling}
\label{decoupling}

In this appendix we decouple Eq.~(\ref{equations12m}) in the dodecahedron 
arrangement.

Let us set $\bmit \eta$ an 11--vector whose components are
\begin{equation}
\begin{array}{rl}
\eta_i & = f_i \qquad\qquad i \in \{1,\ldots,5\} \\
\eta_{i+5} & = q_i \qquad\qquad i \in \{1,\ldots,6\}
\end{array}
\end{equation}
With this position the Lagrangian of the system is written
\begin{equation}
L = \frac{1}{2} T_{ij} \dot \eta^i \dot \eta^j -
\frac{1}{2} V_{ij} \eta^i \eta^j
\end{equation}
where
\begin{equation}
T_{ij} = M \left (
\begin{array}{cc}
(1+3 \alpha^2 \mu/2 \pi) \bmit{\cal I}_5 & \alpha \mu \bmit p \\
\alpha \mu \bmit p^t & \mu \bmit{\cal I}_6
\end{array} \right ) = M t_{ij}
\label{tij}
\end{equation}
is the kinetic energy matrix, while
\begin{equation}
V_{ij} = M \omega_0^2 \left (
\begin{array}{cc}
\bmit{\cal I}_5 & 0 \\
0 & \mu \bmit{\cal I}_6
\end{array} \right ) = M \omega_0^2 v_{ij}
\label{vij}
\end{equation}
is the potential energy matrix. In~(\ref{tij}) and~(\ref{vij}) 
we have put $\mu = m/M$. Motion equation is 
\begin{equation}
T_{ij} \ddot \eta^j + V_{ij} \eta^j = 0
\label{eqeta}
\end{equation}
To decouple the motion, the eigenvalues and eigenvectors of the matrix
\begin{equation}
(V_{ij} - \omega^2 T_{ij}) = 
M \omega_0^2 (v_{ij} - \lambda^2 t_{ij}) 
\label{vt}
\end{equation}
must be found (see~\cite[Cap. 6]{gold80}). In~(\ref{vt}) we set 
$\lambda = \omega/\omega_0$.
Putting equal to zero the determinant of matrix~(\ref{vt}) one obtains
\begin{equation}
\det (V_{ij} - \omega^2 T_{ij}) = (M \omega_0^2)^{11} \mu^6 
(1-\lambda^2) \left [ \lambda^4 - 2 \left (
1 + \frac{3 \alpha^2}{4 \pi} \mu \right ) \lambda^2 + 1
\right ]^5 = 0
\label{det}
\end{equation}
The three distinct eigenvalues are therefore
\begin{eqnarray}
\lambda_{1,2,3,4,5} &=& \lambda_- = 
1 - \frac{1}{2} \sqrt{\frac{3 \alpha^2}{2 \pi} \mu} \nonumber \\
\lambda_6 &=& \lambda = 1 \label{lambda} \\
\lambda_{7,8,9,10,11} &=& \lambda_+ = 
1 + \frac{1}{2} \sqrt{\frac{3 \alpha^2}{2 \pi} \mu} \nonumber
\end{eqnarray}
To look for eigenvectors of~(\ref{vt}) is the same as to find the
matrix $\bmit A$ which simultaneously diagonalizes the matrix of 
kinetic and potential energy $\bmit T$~(\ref{tij}) and
$\bmit V$~(\ref{vij}). 
This matrix is found to be
\begin{equation}
\bmit A = \sqrt{\frac{1}{2 M}}\, \left (
\begin{array}{ccc}
\lambda_- \bmit{\cal I}_5 & \bmit 0_{(5\times 1)} &
\lambda_+ \bmit{\cal I}_5 \\
\ & \ & \ \\
\sqrt{2 \pi/3 \mu} \lambda_- \bmit p^t & 
\bmit 1_{(6\times 1)}/\sqrt{3 \mu} & 
- \sqrt{2 \pi/3 \mu} \lambda_+ \bmit p^t
\end{array} \right )
\label{aaa}
\end{equation}
where we have introduced notation $\bmit k_{(L \times N)}$ to indicate 
a matrix with dimensions $L \times N$ whose elements are all equal to 
$k$. $\bmit A$ is also the transition matrix between the generalized 
coupled coordinates $\eta_i$ to the normal coordinates $\zeta_i$
\begin{equation}
\eta_i = \sum_j{A_{ij} \zeta_j}
\end{equation}
Let us now consider the forced system. Let $\bmit F = (F_j)$ be the 
generalized forces corresponding to the generalized coordinates
$\bmit \eta = (\eta_j)$. Then, generalized forces $\bmit Q = (Q_i)$ 
for the normal coordinates are
\begin{equation}
\bmit Q = \bmit A^t \bmit F
\end{equation}
If the external forces act only on the sphere modes then 
(see~(\ref{equations12m}))
\begin{equation}
\begin{array}{rll}
F_i =& M g_i \qquad\qquad & i \in \{1,\ldots,5\} \\
F_i =& 0 & i \in \{6,\ldots,11\}
\end{array}
\end{equation}
Therefore
\begin{equation}
\bmit Q = \sqrt{\frac{M}{2}}\, \left (
\begin{array}{c}
\lambda_- \bmit g \\
0 \\
\lambda_+ \bmit g
\end{array} \right )
\end{equation}
If we put
\begin{eqnarray}
\bmit \zeta_- &=& (\zeta_i) \qquad\qquad i \in \{1,\ldots,5\}
\nonumber \\
\zeta &=& \zeta_6 \\
\bmit \zeta_+ &=& (\zeta_i) \qquad\qquad i \in \{7,\ldots,11\}
\nonumber 
\end{eqnarray}
then eqs.~(\ref{equationsnm}) and (\ref{conv}) are immediately found.

\section{Calculation of $\bmit{\lowercase{g}}$}
\label{force}

In this appendix we show how to find the components of vector
$\bmit g_n$ (see Sec.~\ref{gravity}). First the amplitudes 
$R'\!_{nlm'}(t)$, defined in~(\ref{gravacc}) must be calculated.
With the given gravitational signal~(\ref{gravwave}) one has that the 
only coefficients which does not vanish in $O'$ are (cfr. 
with~\cite{wagpai77}):
\begin{eqnarray}
R'_{n22}(t) & = & - \sqrt{\frac{32 \pi}{15}} R (\alpha_{n2} +
3 \beta_{n2}) E_R(t) \nonumber \\
R'_{n2-2}(t) & = & - \sqrt{\frac{32 \pi}{15}} R (\alpha_{n2} +
3 \beta_{n2}) E_L(t) \label{Rdz} \\
R'_{n20}(t) & = & - \sqrt{\frac{16 \pi}{15}} R (\alpha_{n2} +
3 \beta_{n2}) E_l(t) \nonumber 
\end{eqnarray}
where, using the same notation as~\cite{wagpai77}, we have
\begin{equation}
\alpha_{n2} = (M R)^{-1} \int a_{nl}(r) \rho r^3\, dr
\qquad\qquad
\beta_{n2} = M^{-1} \int b_{nl}(r) \rho r^2\, dr
\end{equation}
Now $G_{nml}(t)$ defined in~(\ref{gvsrp}), that is to say the 
coefficients of the gravitational wave decomposition in $O$ are to be
evaluated. Because of~(\ref{Rdz}) one has
\begin{equation}
G_{nml}(t) = \delta_{l2} \left (
{\cal D}^{(2)}_{m2} R'_{n22}(t) + 
{\cal D}^{(2)}_{m-2} R'_{n2-2}(t) +
{\cal D}^{(2)}_{m0} R'_{n20}(t)
\right )
\end{equation}
Following~\cite{edm57,wagpai77} one has
\begin{equation}
{\cal D}^{(2)}_{mm'}(H,\delta) = 
\exp{[i m H]} d^{(2)}_{mm'}(\delta + \pi/2)
\label{calD}
\end{equation}
where $d^{(2)}_{mm'}(\beta)$ ($\beta = \delta + \pi/2$) 
are given by (see~\cite{edm57})
\begin{equation}
\bmit d^{(2)}(\beta) = \left (
\begin{array}{ccccc}
d_{22}(\beta) & -d_{12}(\beta) & d_{02}(\beta) & -d_{-12}(\beta) & 
d_{-22}(\beta) \\
d_{12}(\beta) & d_{11}(\beta) & d_{10}(\beta) & d_{1-1}(\beta) & 
-d_{-12}(\beta) \\
d_{02}(\beta) & -d_{10}(\beta) & d_{00}(\beta) & d_{10}(\beta) & 
d_{02}(\beta) \\
d_{-12}(\beta) & d_{1-1}(\beta) & -d_{10}(\beta) & d_{11}(\beta) & 
-d_{12}(\beta) \\
d_{-22}(\beta) & d_{-12}(\beta) & d_{02}(\beta) & d_{12}(\beta) & 
d_{22}(\beta) 
\end{array} \right )
\end{equation}
where
\begin{eqnarray*}
d_{22}(\delta+\pi/2) &=& 3/8 - 1/2 \sin \delta - 1/8 \cos 2\delta
\\
d_{-22}(\delta+\pi/2) &=& 3/8 + 1/2 \sin \delta - 1/8 \cos 2\delta
\\
d_{12}(\delta+\pi/2) &=& - 1/2 \cos \delta + 1/4 \sin 2\delta
\\
d_{-12}(\delta+\pi/2) &=& - 1/2 \cos \delta - 1/4 \sin 2\delta
\\
d_{11}(\delta+\pi/2) &=& - 1/2 \sin \delta - 1/2 \cos 2\delta
\\
d_{1-1}(\delta+\pi/2) &=& - 1/2 \sin \delta + 1/2 \cos 2\delta
\\
d_{02}(\delta+\pi/2) &=& \sqrt{3/32}\ (1+ \cos 2\delta)
\\
d_{10}(\delta+\pi/2) &=& - \sqrt{3/8} \sin 2\delta
\\
d_{00}(\delta+\pi/2) &=& (1 - 3 \cos 2\delta)/4
\end{eqnarray*}
Setting $\gamma_n = - \sqrt{32 \pi/15\, }\, R (\alpha_{n2} + 
3 \beta_{n2})$, the irreducible tensor of second rank $\bmit G_{n}$, 
whose components are $G_{n2m}$ is written
\begin{equation}
\bmit G_{n} = \gamma_n \left[ 
E_R(t) \left (
\begin{array}{c}
d_{22} e^{i 2 H} \\
d_{12} e^{i H} \\
d_{02} \\
d_{-12} e^{-i H} \\
d_{-22} e^{-i 2 H} 
\end{array} \right ) +
E_L(t) \left (
\begin{array}{c}
d_{-22} e^{i 2 H} \\
-d_{-12} e^{i H} \\
d_{02} \\
-d_{12} e^{-i H} \\
d_{22} e^{-i 2 H}
\end{array} \right ) +
\frac{E_l(t)}{\sqrt{2}} \left (
\begin{array}{c}
d_{02} e^{i 2 H} \\
d_{10} e^{i H} \\
d_{00} \\
-d_{10} e^{-i H} \\
d_{02} e^{-i 2 H}
\end{array} \right )
\right ]
\label{VCG}
\end{equation}
Applying to $\bmit G_{n}$ the operator $\bmit U^{(2)}$ defined 
in~(\ref{unitaryu}) we get the real vector $\bmit g_n$ whose 
components are given by~(\ref{vectorg}).

\section{inversion of system~(\lowercase{\ref{vectorg})}}
\label{inversion}

After manipulation of system~(\ref{vectorg}) with 
definition~(\ref{vectora}) we find the following equation for
$x=\tan{H/2}$:
\begin{equation}
c_{(0)} + 2\,c_{(1)}\,x + c_{(2)}\,x^2 + 4\,c_{(3)}\,x^3 -
c_{(2)}\,x^4 + 2\,c_{(1)}\,x^5 - c_{(0)}\,x^6 = 0
\label{tai}
\end{equation}
where
\begin{eqnarray}
c_{(0)} &=& a_2^2\,a_3 - 
     {\sqrt{3}}\,a_1\,a_2\,a_4 - a_3\,{{a_4}^2} + 
     a_2\,a_4\,a_5 
\nonumber \\
c_{(1)} &=&  {{a_2}^3} - 2\,a_2\,{{a_3}^2} + 
\nonumber \\
&+& {\sqrt{3}}\,a_1\,a_3\,a_4 + 
     a_2\,{{a_4}^2} + 
     2\,{\sqrt{3}}\,a_1\,a_2\,a_5 + 
     3\,a_3\,a_4\,a_5 - 2\,a_2\,{{a_5}^2}  
\nonumber \\ 
c_{(2)} &=& -11\,{{a_2}^2}\,a_3 + 4\,{{a_3}^3} + 
\nonumber \\
&+& 7\,{\sqrt{3}}\,a_1\,a_2\,a_4 + 
     7\,a_3\,{{a_4}^2} - 
     8\,{\sqrt{3}}\,a_1\,a_3\,a_5 - 
     15\,a_2\,a_4\,a_5 - 8\,a_3\,{{a_5}^2} 
\nonumber \\ 
c_{(3)} &=& -{{a_2}^3} + 4\,a_2\,{{a_3}^2} - 
\nonumber \\
&-& 3\,{\sqrt{3}}\,a_1\,a_3\,a_4 - 
     3\,a_2\,{{a_4}^2} - 
     2\,{\sqrt{3}}\,a_1\,a_2\,a_5 - 
     5\,a_3\,a_4\,a_5 + 2\,a_2\,{{a_5}^2} 
\nonumber 
\end{eqnarray}
Declination angle $\delta$ is found from
\begin{equation}
\tan \delta =
\frac{a_4 \cos 2 H - a_5 \sin 2 H}
{a_2 \cos H - a_3 \sin H}
\label{tgd}
\end{equation}
while the three polarization amplitudes are given by
\begin{eqnarray}
E_\times &=& (a_4 \cos 2 H - a_5 \sin 2 H) \sin \delta +
(a_2 \cos H - a_3 \sin H) \cos \delta
\label{eper} \\
\left (E_+ - \sqrt{3} E_l \right ) &=&
2 (a_3 \cos H + a_2 \sin H) \sin 2 \delta -
(a_5 \cos 2 H + a_4 \sin 2 H -\sqrt{3} a_1) \cos 2 \delta
\label{epmel} \\
\left (E_+ +\frac{1}{\sqrt{3}} E_l \right ) &=& 
a_5 \cos 2 H + a_4 \sin 2 H + \frac{a_1}{\sqrt{3}}
\label{eppel}
\end{eqnarray}


\begin{figure}
\caption{Simulation over 1000 attempts of reconstructed position of a 
source located at hour angle $H=1.0\,rad$ and declination
$\delta=0.7\,rad$. Amplitude signal to noise ratio is 10. The incoming 
gravitational wave is assumed linearly polarized with 
$E_\times/E_+=3/4$.}
\label{fig1}
\end{figure}

\begin{figure}
\caption{The same as Fig.~\ref{fig1} with $H=1.0\,rad$,
$\delta=1.3\,rad$ and $E_\times/E_+=4/3$.}
\label{fig2}
\end{figure}

\newpage

\begin{table}
\caption{Reconstruction over 1000 attempts of gravitational signals 
coming from sources placed at hour angle $H$ and declination $\delta$.
$\Delta\Omega$ is the angular resolution in steradians; 
$E$ is the amplitude of the incoming waves in units of the $SNR$; 
$E_+$ and $E_\times$ are the polarization amplitudes; $E_l$ is the 
eventual longitudinal component of the signal, which is used as a veto 
against spurious events.\\
\label{tab1}}
\begin{tabular}{c|c|c|c|c|} 
\ & \multicolumn{2}{c|}{First case} & 
\multicolumn{2}{c|}{Second case} \\ \hline
\ & Original data & Reconstruction & Original data & Reconstruction \\ 
\hline
$\delta\,(rad)$ & $0.7$ & $0.7\pm 0.1$ & $1.3$ & $1.3\pm 0.1$ \\ \hline
$H\,(rad)$ & $1.0$ & $1.0\pm 0.15$ & $1.0$ & $1.0\pm 0.5$  \\ \hline
$\Delta \Omega\,(sterad)$ & \ & $1\times 10^{-2}$ & \ & $1\times 10^{-2}$ 
\\ \hline
$E\,(SNR)$ & $10$ & $10\pm 1$ & $10$ & $10\pm 1$ \\ \hline
$E_+$ & $8$ & $8\pm 1.5$ & $6$ & $4\pm 5$ \\ \hline
$E_\times$ & $6$ & $6\pm 2$ & $8$ & $6.0\pm 4.5$  \\ \hline
$E_l/E$ & $0$ & $0.0\pm 0.1$ & $0$ & $0.0\pm 0.1$ \\ \hline
\end{tabular}
\end{table}

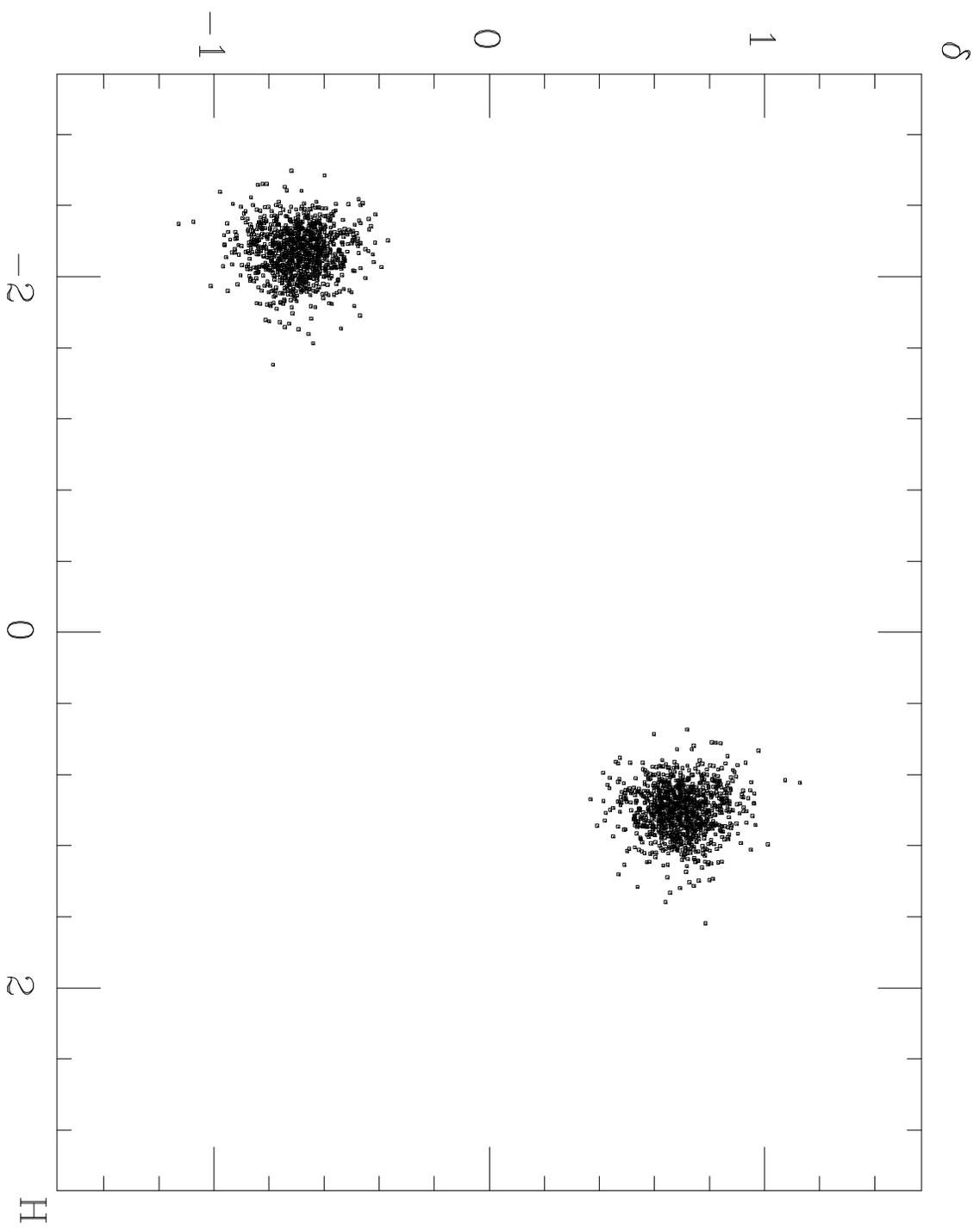
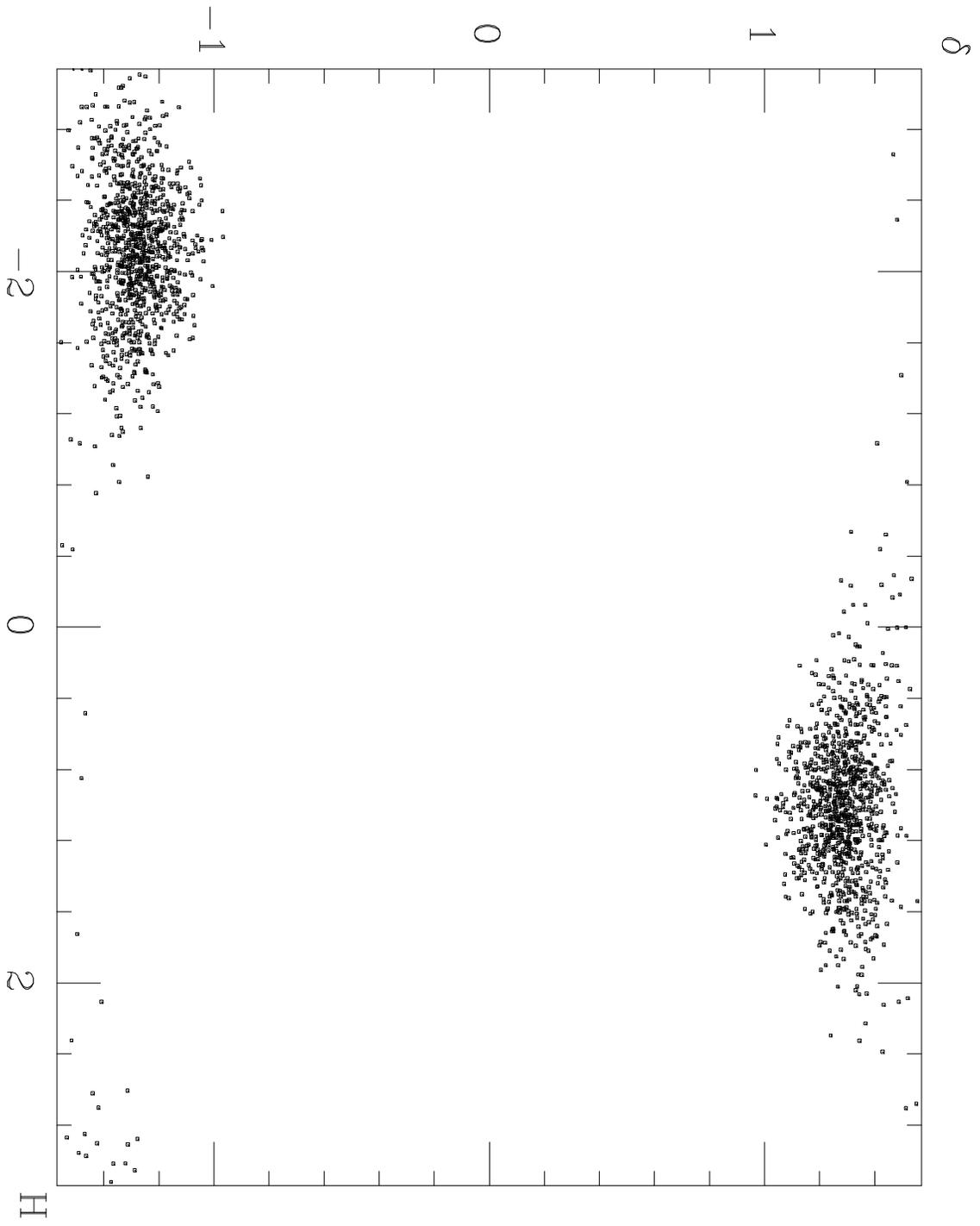
\end{document}